\documentclass{fundam}

\usepackage{amsmath}
 \usepackage{flagderiv}

\usepackage{prooftree}
\usepackage{amsfonts}
\usepackage{epsfig}
\usepackage[english]{babel}
\usepackage{graphicx}

\usepackage{macros}

   \usepackage[colorlinks = true, linkcolor = black, urlcolor = black, citecolor = black, anchorcolor = black]{hyperref}

\begin{document}
	
\setcounter{page}{313}
\publyear{22}
\papernumber{2112}
\volume{185}
\issue{4}

   \finalVersionForARXIV

\title{Characteristics of de Bruijn's early proof checker Automath}

\author{Herman Geuvers\thanks{Address of correspondence: H.\ Geuvers, Faculty of Science, Radboud University,
                                 PO Box 9010, 6500 GL Nijmegen, The Netherlands. \newline \newline
          \vspace*{-6mm}{\scriptsize{Received June 2021; \ accepted April  2022.}}}
          \\
  Radboud University Nijmegen and \\
  Eindhoven University of Technology\\
  Nijmegen, The Netherlands\\
   herman@cs.ru.nl
  \and
  Rob Nederpelt\\
Eindhoven University of Technology \\
 Nijmegen, The Netherlands
}

\date{}

\runninghead{H.\ Geuvers and R.\ Nederpelt}{Characteristics of de Bruijn's Automath}

\maketitle

\begin{abstract}
The `mathematical language' Automath, conceived by N.G. de Bruijn in 1968, was the first theorem prover actually working and was used for checking many specimina of mathematical content. Its goals and syntactic ideas inspired Th.\ Coquand and G.\ Huet to develop the calculus of constructions, CC (\cite{CoqHue}), which was one of the first widely used interactive theorem provers and forms the basis for the widely used Coq system (\citealp{Coq}).

The original syntax of Automath (\cite{deB1970,NedGeuDeV}) is not easy to grasp. Yet, it is essentially based on a derivation system that is similar to the Calculus of Constructions (`CC'). The relation between the Automath syntax and CC has not yet been sufficiently described, although there are many references in the type theory community to Automath.

In this paper we focus on the backgrounds and on some uncommon aspects of the syntax of Automath. We expose the fundamental aspects of a `generic' Automath system, encapsulating the most common versions of Automath. We present this generic Automath system in a modern syntactic frame. The obtained system makes use of $\lambda$D, a direct extension of CC with definitions described in \cite{NedGeu}.
\end{abstract}

\section{Introduction}\label{Introd}

Around 1967, N.G.\ de Bruijn started thinking about a formal checker for mathematical proofs. He defined several Automath languages, of which AUT68 and AUT-QE (see \citealp{deB1980}) are the most used ones. They differ in `strength' as to the abstraction facilities, but have a common structure. Below, we concentrate on the general aspects of the Automath languages, without differentiating between the different `dialects'. Therefore we speak about `Automath' only and discuss its structure in a {\em generic\/} manner.

The purpose of this paper is to give an impression of what Automath
‘is’ and discuss in passing what de Bruijn intended with this
‘mathematical language’. In Section~\ref{AutFor}, we describe what the
features of Automath are, as compared to other modern type theories
and proof assistants based on type theory. In particular we focus on
the `book format' of Automath, with the idea that a formal
mathematical text is built up incrementally, line by line, where each
new line has to be correct in terms of the already existing part of
the book, and thereby one creates a correct Automath book. In
Section~\ref{AutMod} we define formally what this correctness means in
type-theoretic terms. In Section~\ref{Conc} we discuss de Bruijn's basic ideas about
Automath and we summarise how Automath relates to modern type theories.

\section{The Automath format}\label{AutFor}

To give an idea of the original syntax of Automath,
we reproduce a short fragment from the beginning of the formalization of Landau's Grundlagen (\citealp{Lan}), written by L.S. van Benthem Jutting (\citealp{Jutting}) in Figure~\ref{AutForLan}. In Section~\ref{AutForInt} we give a translation of this text in (more or less) common mathematical phrasings.

\weg{
\begin{figure}[h]
\begin{tabular}{r  c  l  c  l  l  l}
& $\ast$ & $A$ & $ := $ & --- & ; & {\it PROP}\\
$A$ & $\ast$ & $B$ & $ := $ & --- & ; & {\it PROP}\\
$B$ & $\ast$ & $\IMP$ & $ := $ & $[X,A]B$ & ; & {\it PROP}\\
$B$ & $\ast$ & $C$ & $ := $ & --- & ; & {\it PROP}\\
$C$ & $\ast$ & $I$ & $ := $ & --- & ; & $\IMP(A,B)$\\
$I$ & $\ast$ & $J$ & $ := $ & --- & ; & $\IMP(B,C)$\\
$J$ & $\ast$ & $\TRIMP$ & $ := $ & $[X,A]<<X>I>J$ & ; & $\IMP(A,C)$\\
& $\ast$ & $\CON$ & $ := $ & \PN & ; & {\it PROP}\\
$A$ & $\ast$ & $\NOT$ & $ := $ & $\IMP(\CON)$ & ; & {\it PROP}\\
$A$ & $\ast$ & $\WEL$ & $ := $ & $\NOT(\NOT(A))$ & ; & {\it PROP}\\
$A$ & $\ast$ & $A1$ & $ := $ & --- & ; & $A$\\
$A1$ & $\ast$ & $\WELI$ & $ := $ & $[X,\NOT(A)]<A1>X$ & ; & $\WEL(A)$\\
$A$ & $\ast$ & $W$ & $ := $ & --- & ; & $\WEL(A)$\\
$W$ & $\ast$ & $\ET$ & $ := $ & \PN & ; & $A$\\
$A$ & $\ast$ & $C1$ & $ := $ & --- & ; & $\CON$\\
$C1$ & $\ast$ & $\CONE$ & $ := $ & $\ET([X,\NOT(A)]C1)$ & ; & $A$
\end{tabular}
\caption{An example of an Automath text}
\label{AutForLan}
\end{figure}
}

\begin{figure}[h]
\centering
  \includegraphics[width=0.9\linewidth]{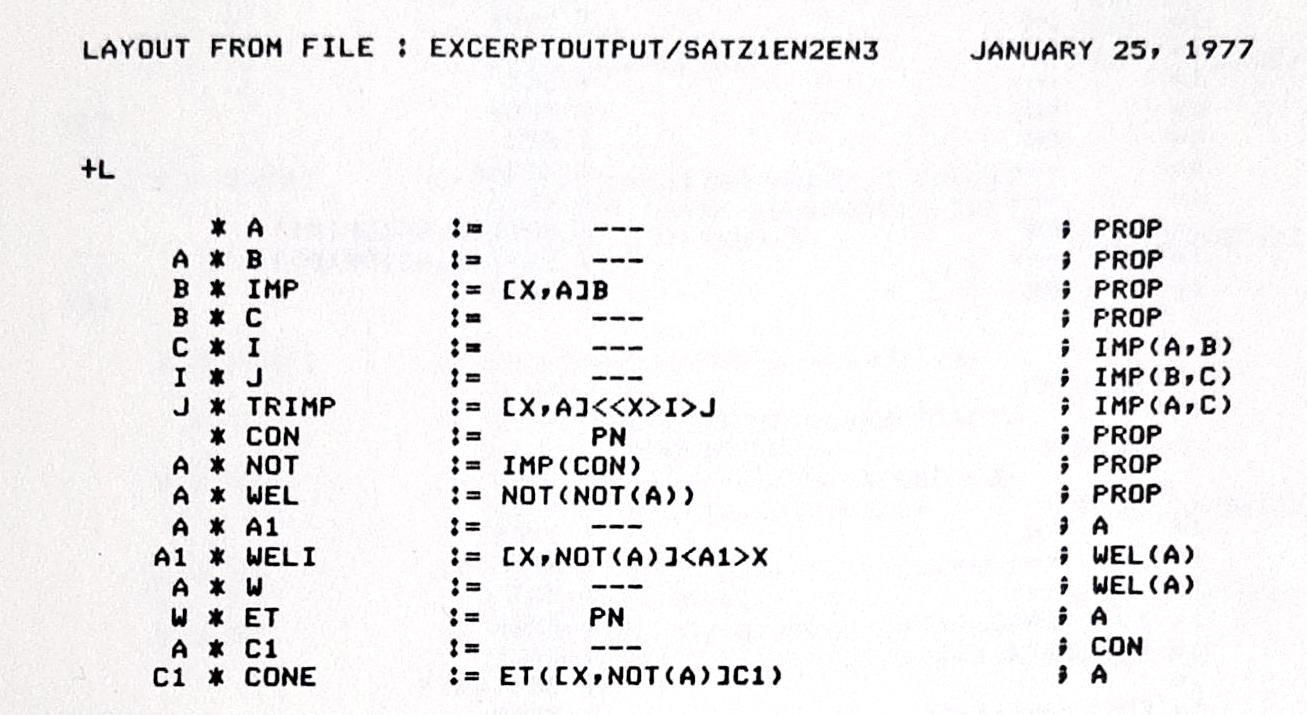}\vspace*{-1mm}
  \caption{An example of an Automath text}
\label{AutForLan}
\end{figure}

As can be seen from this example, an Automath text consists of consecutive lines, each containing four expressions. These lines can have either of the three forms $(i)$, $(ii)$ or $(iii)$ of Figure~\ref{AutForThr}.

\begin{figure}[!h]
\centering
\begin{tabular}{c||clclclc||l}
& indicator &  & identifier & & content & & category & \\
\hline \hline
\rule{0em}{1.1em}
 \hspace{-1.5ex} $(i) $\hspace{-1ex} & $\epsilon / x$ & \hspace{-1ex} $\ast$ \hspace{-1em}  & $z$ & \hspace{-1em} $:=$ \hspace{-1em}  & --- & \hspace{-1ex} $:$ \hspace{-1em}  & $A$ & 
 assumption line\\
\hline
\rule{0em}{1.1em}
 \hspace{-1.5ex} $(ii)$ \hspace{-1ex} & $\epsilon / x$ & \hspace{-1ex} $\ast$ \hspace{-1em} & $c$ & \hspace{-1em} $:=$ \hspace{-1em} & $M$ & \hspace{-1ex} $:$ \hspace{-1em}  & $A$ & 
 definition line\\
\hline
\rule{0em}{1.1em}
\hspace{-1.5ex} $(iii)$ \hspace{-1ex} & $\epsilon / x$ & \hspace{-1ex} $\ast$ \hspace{-1em} & $c$ & \hspace{-1em} $:=$\hspace{-1em}  & \PN & \hspace{-1ex} $:$  \hspace{-1em} & $A$ & 
 primitive line\\
\hline
\end{tabular}
\caption{The three species of Automath lines}
\label{threespecies}
\end{figure}\label{AutForThr}

In this scheme, $\epsilon$ (denoting the empty expression) is a special symbol, $x$ and $z$ represent variables, $c$ denotes a constant and $M$ and $A$ are expressions; also \PN\ acts as a special symbol. The symbols $\ast$, $:=$ and $:$ between the expressions are separators. \footnote{Instead of the semicolon `;' used in Automath, we take the colon~`:' -- as is common nowadays -- for expressing the relation between a content and a category (or `type').}

\smallskip
The {\it first\/} entry of all Automath-lines, the {\it indicator\/} or {\it context indicator\/}, is the special symbol $\epsilon$ or a variable (here with the name $x$); the indicator acts to establish the context of the line.

In the {\it second\/} entry, an {\it identifier\/} (or {\it name\/}) is introduced. This can be either a {\it variable\/} (such as $z$, in assumption lines) or a {\it constant\/} (say $c$, in definition lines or primitive lines).

The {\it third\/} entry is called the {\it content\/}.\footnote{The content is called the {\it definition\/} in the original Automath paper. We use the word {\it content\/} to avoid confusion with the general meaning of the word `definition'.}
A content is the symbol `---' when the name is a variable (in assumption lines), and an expression $M$ or the symbol \PN\ in the two other cases.

\vspace{0.3mm}
The {\it fourth\/} entry, the {\it category\/}, contains an expression $A$.\footnote{We shall often use the word `type' instead of `category', in order to connect Automath with more common terminology in type theory.}

\medskip
The {\it meaning\/} of the three kinds of lines can be described shortly as follows: an assumption line contains a {\it context extension\/}; a definition line introduces a {\it definition\/} and a primitive line introduces a {\it postulate\/} (both in a given context).

\medskip
More explicitly:

\smallskip
$(i)$ {\it assumption line\/}: in the empty context (if the indicator is $\epsilon$) or in a context having indicator $x$, the {\em assumption\/} is made that newly introduced variable $z$ has category $A$; or, in type-theoretic terms, we extend a given context (being empty, or having indicator $x$) with $z : A$, the {\it declaration\/} of variable $z$ of {\it type\/} $A$;

\smallskip
$(ii)$ {\it definition line\/}: in the empty context ($\epsilon$) or in a context having indicator $x$, constant $c$ is {\em defined\/} as $M$ of category $A$;

\smallskip
$(iii)$ {\it primitive line\/}: in the empty context ($\epsilon$) or in a context having indicator $x$, constant $c$ is {\em postulated\/} as a primitive notion  (`\PN') of category $A$.

\medskip
In order to clarify these notions, we start with a discussion of contexts in general (Section~\ref{AutForCon}) and, in particular, in Automath-style (\ref{AutForAss}). This is followed by the description of how formalised definitions (\ref{AutForDef}) and postulates (\ref{AutForPri}) are dealt with in Automath.

\subsection{Contexts}\label{AutForCon}

A (type-theoretical) context consists of a list of {\em declarations\/} of so-called {\em declared variables\/} $x_1$, \ldots, $x_n$, together with their {\em types\/} $A_1$, \ldots, $A_n$, respectively. In type theory, such a list is generally written as
$x_1 : A_1$, $x_2 : A_2$, \ldots, $x_n : A_n$,
or, shortly, as $\overline{x} : \overline{A}$.

Before we describe how contexts are treated in Automath (see Section~\ref{AutForAss}), we devote some space to two  different manners in which contexts are expressed. For that we first give a mathematical example of how contexts are introduced, followed by the same example denoted in flag-style (or Fitch-style; see below).

\smallskip
In mathematics and logic, contexts are omnipresent. See the following example, which illustrates their use:

\begin{example}[Contexts in mathematics]\label{AutForConExa}

{\it
\smallskip
\noindent {\tt LEMMA 5.2 \hspace{0.2em}(a) Let $A$, $B$ and $C$ be subsets of a given set $S$. \hspace{-0.6em}Assume

$A \subseteq B$ and $B \subseteq C$. \hspace{-0.6em}Then also $A \subseteq C$.

(b) Let, moreover, $C \subseteq A$. \hspace{-0.6em}Then $A = B$.

(c) $A \cap B \subseteq A \cup B$.

(d) Let $A = B$. \hspace{-0.6em}Then $A \cap B = A \cup B$.}
}
\end{example}

\begin{proof}
Proof of (a):  Let $x$ be an element of $A$. Then $x \in B$ so also $x \in C$.
\end{proof}

The various contexts in this example can be expressed as follows
in a typetheoretical style, using the propositions-as-types isomorphism.

\begin{remark}\label{AutForConNot} We only want to exemplify the notion `context'. The symbol ${\tt set}$ represents the `universe' of sets. We do not explain why we introduce a `universe' $S$. We identify sets in the universe by means of the power set of $S$: a set is an inhabitant of type ${\cal P}(S)$. Moreover, we freely employ infix notation, although this is against the Automath conventions. We also deviate a bit from the Automath precision by not mentioning the $S$ when using the symbols $\subseteq$, $=$, $\cap$, $\cup$ and $\in$. For details, see~\citealp{NedGeu}.
\end{remark}

\noindent{\tt Context of part (a):}

$S : {\tt set},~A : {\cal P}(S),~B : {\cal P}(S),~C : {\cal P}(S),~u : A \subseteq B,~v : B \subseteq C$.

\noindent{\tt Context of part (b):}

 $S : {\tt set},~A : {\cal P}(S),~B : {\cal P}(S),~C : {\cal P}(S),~u : A \subseteq B,~v : B \subseteq C,~w : C \subseteq A$.

\noindent{\tt Context of part (c):}

$S : {\tt set},~A : {\cal P}(S),~B : {\cal P}(S)$.

\noindent{\tt Context of part (d):}

 $S : {\tt set},~A : {\cal P}(S),~B : {\cal P}(S),~y : (A = B)$.

\smallskip

\noindent{\tt Context of PROOF of part (a):}

 $S : {\tt set},~A : {\cal P}(S),~B : {\cal P}(S),~C : {\cal P}(S),~u : A \subseteq B,~v : B \subseteq C,~x : S,$

\noindent~$~z : (x \in A)$.

\medskip
It will be clear that the type-theoretical style requires a frequent repetition of initial parts of contexts, in particular for the tree-like presentation as is known from proof theory. Such repetitions are to a great extent avoided in the linear presentation as is common in mathematics texts (see the Lemma above).

In formal versions of mathematics, the duplication of context elements is unpleasant. One way out is to use a {\em Fitch-style\/} notation\footnote{This presentation of natural deduction already appears in the work of Stanis{\l}aw Ja\'{s}kowski (\citealp{Jas}).} (see \citealp{Fit}). Fitch introduced this notation in his logical course, with the clear intention to condense equal parts of contexts. He uses a linear presentation to derive logical proofs, embedded in nested rectangles to indicate the span of an assumption: each initial entry of a rectangle is supposed to be an assumption -- i.e., a context entry -- in all (and only) the statements within that rectangle, so also in the deeper nested ones. This gives a {\em block structure\/} to texts, as is well-known from programming languages such as Algol. For an overview of how this notation is used in systems of natural deduction, see \citealp{Sta}.

A variant of the Fitch-notation is the so-called {\em flag notation\/}, introduced in the 1970's, independently from Fitch, by R.P.\ Nederpelt and N.G.\ de Bruijn. It uses a {\em flag\/} for each initial entry of a block (acting as an assumption), and a {\em flag pole\/} to indicate the range of that assumption (and thus the length

 \setlength{\derivskip}{4pt}
\begin{figure}[!h]
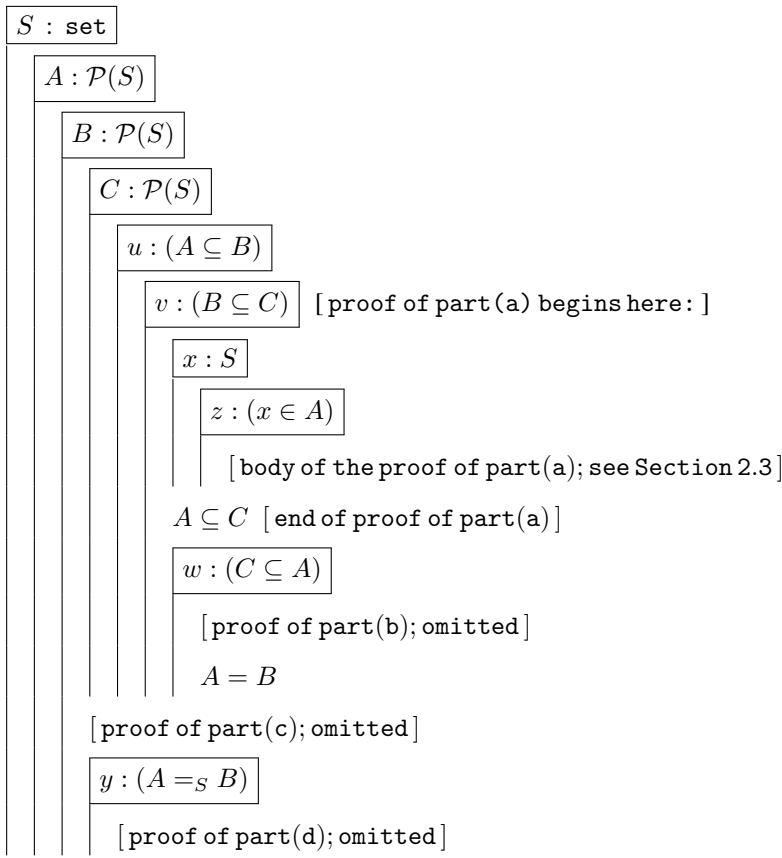
\small
\centering
\begin{flagderiv}
\assume*{}{S \,:\, {\tt set} }{}
\assume*{}{A : {\cal P}(S)}{}
\assume*{}{B : {\cal P}(S)}{}
\assume*{}{C : {\cal P}(S)}{}
\assume*{}{u : (A \subseteq B)}{}
\assume*{}{v : (B \subseteq C)}{\hspace{-7cm}[\,{\tt proof\,of\,part\!\,(a)\,begins\,here:}\,]}
\assume*{}{x : S}{}
\assume*{}{z : (x \in A)}{}
\step*{}{[\,{\tt body\,of\,the\,proof\,of\,part\!\,(a); see\,Section\,\ref{AutForDef}}\,]}{}
\done
\done
\step*{}{A \subseteq C~~ [\,{\tt end\,of\,proof\,of\,part\!\,(a)}\,]}{}
\assume*{}{w : (C \subseteq A)}{}
\step*{}{[\,{\tt proof\,of\,part\!\,(b); omitted}\,]}{}
\step*{}{A = B}{}
\done
\done
\done
\done
\step*{}{[\,{\tt proof\,of\,part\!\,(c); omitted}\,]}{}
\assume*{}{y : (A =_S B)}{}
\step*{}{[\,{\tt proof\,of\,part\!\,(d); omitted}\,]}{}
\end{flagderiv}
\caption{Example of a derivation in flag style}
\label{AutForFla}\vspace*{-3mm}
\end{figure}

 \eject
 \noindent of the block). In Figure~\ref{AutForFla} we express the structure of the above example lemma in flag format, in order to show how it works. (See also \citealp{NedGeu}.)

It is obvious that the flag- (or Fitch-)notation allows {\em sharing\/} of declarations: a comparison with the series of example contexts presented above, shows that one flag can cover several context items, so repetitions have been avoided.

In the example we illustrated that assumptions phrased in a lemma or theorem can be translated as newly introduced context items, that can be withdrawn later. But also the application of {\em logical\/} rules may lead to context changes (either {\em context extension\/} or {\em context shortening\/}). This can be seen, again, in the previous example, where the assumptions $x : S$ and
$z : (x \in A)$ are introduced in order to obtain the logical result $A \subseteq C$ (i.e., $\forall_{x : S}(x \in A \Rar x \in C)$)as the final result of part~(a). We come back to this in Section~\ref{AutForDef}.

\subsection{Assumption lines in Automath}\label{AutForAss}

Automath goes still further in avoiding repetitions of context assumptions. De Bruijn's idea was to use a handy {\em naming convention\/} for contexts: each context is represented by its {\em last\/} declared variable, acting as the {\it indicator\/} of the complete context. So the four contexts of parts~(a) to (d) of the Lemma given in the previous section (cf.\ Figure~\ref{AutForFla}), are simply referred to as $v$, $w$, $B$ and $y$, respectively.

\medskip
We shall explain below why this is advantageous. But firstly we note that this naming only works smoothly if we agree never to introduce the name of a declared variable more than once. This {\it unique naming convention\/} for context variables is, of course, rather restrictive and cannot be maintained long. However, in theory (and in this paper) we stick to it, noting that de Bruijn found a practical way out in Automath by using a so-called {\em paragraph system\/} (see \citealp{Jutting}, p.\ 78 or \citealp{Zan}, Section~11), thus securing the desired unicity of names for context variables, and for defined constants, as well.

\begin{remark}\label{AutForNott} (1) In Automath all defined constants are supposed to be different, as well, and taken from another set than variables. This has the same drawbacks regarding unicity, which can be overcome similarly as with context variables.

(2) {\it Abstraction variables\/} are named independently of the names of context variables and defined constants. Moreover they do not obey the unique naming convention. For example, the name $x$ in $\lambda x : A . B$ (or in $\Pi x : A . B$) can be multiply used, as long as the ranges of the various x's are not interfering.
\end{remark}

In Automath-style, a single assumption $z : N$ can be added as follows to an already known context named $x$:

\smallskip
\begin{tabular}{l c l c l c l}
$x$ & $\ast$ & $z$ & $:=$ & $\text{---}$ & $:$ & $N$
\end{tabular}

\smallskip
\noindent Such a line is called an {\em assumption line\/}. In case the context to which $z : N$ is added, is the {\em empty context\/} (i.e., if the assumption stands on its own), then the symbol $\epsilon$ is used:

\smallskip
\begin{tabular}{l c l c l c l}
$\epsilon$ & $\ast$ & $z$ & $:=$ & $\text{---}$ & $:$ & $N$
\end{tabular}

\medskip
As an example, we express the contexts belonging to the Lemma parts~(a) to (d) mentioned in the previous section, in Automath format.

\eject
We start with part~(a), in which the context consists of six assumptions:

\smallskip
\begin{tabular}{l c l c l c l}
$\epsilon$ & $\ast$ & $S$ & $:=$ & $\text{---}$ & $:$ & ${\tt set}$\\
$S$ & $\ast$ & $A$ & $:=$ & $\text{---}$ & $:$ & ${\cal P}(S)$\\
$A$ & $\ast$ & $B$ & $:=$ & $\text{---}$ & $:$ & ${\cal P}(S)$\\
$B$ & $\ast$ & $C$ & $:=$ & $\text{---}$ & $:$ & ${\cal P}(S)$\\
$C$ & $\ast$ & $u$ & $:=$ & $\text{---}$ & $:$ & $A \subseteq B$\\
$u$ & $\ast$ & $v$ & $:=$ & $\text{---}$ & $:$ & $B \subseteq C$
\end{tabular}

\bigskip
Now it is easy to express the context for part~(b), which apparently consists of only one assumption (the other six are covered by the indicator $v$):

\smallskip
\begin{tabular}{l c l c l c l}
$v$ & $\ast$ & $w$ & $:=$ & $\text{---}$ & $:$ & $C \subseteq A$
\end{tabular}

\medskip
The context for part~(c) can be described simply by `plugging in' at the identifier $B$, which is introduced in the third assumption of part~(a).
So if we need that context, we just put a $B$ in the `indicator' column.

\smallskip
This is exemplified in the Automath-version of the context for part~(d):

\smallskip
\begin{tabular}{l c l c l c l}
$B$ & $\ast$ & $y$ & $:=$ & $\text{---}$ & $:$ & $A = B$
\end{tabular}

\medskip
The Automath-notation of contexts is still more `economic' than the flag notation. Using flags in Fitch-style, a closed block (i.e., the ending of a flag pole) cannot be reopened. Yet, such a thing is regularly the case in mathematics. For example, one could continue Figure~\ref{AutForFla} with describing another lemma about a triple of sets, $A$, $B$ and $C$. But since the flag of $C$ has been closed already, the only way out is to re-introduce a flag with $C$.

In the Automath-style, this is not needed, even when all flags in Figure~\ref{AutForFla} have been closed and the text in Figure~\ref{AutForLan} is far out of sight. One just puts the $C$ as the indicator of a new line and the context named $C$ is available, again. Otherwise said, the flags of $S$, $A$, $B$ and $C$ have been reopened.

\subsection{Definition lines in Automath}\label{AutForDef}

The possibility to add definitions is essential in mathematics. One generally uses a special format for definitions. For example, previous to the above example called `Lemma~5.2' in Example~\ref{AutForConExa} there will have appeared a definition for inclusion of sets, for example:

\begin{definition}
 Let $A$ and $B$ be sets. One says that $A$ is included in $B$ (denoted $A \subseteq B$) if for all $x \in A$ it holds that $x \in B$.
\end{definition}

Note that this definition needs a {\it context\/} again, namely $S : {\tt set},~A : {\cal P}(S)$, $B : {\cal P}(S)$. This is the context named $B$ in the previous section.

\smallskip

In Automath, we use a line of the form~$(ii)$ for stating a definition. Using the context named $B$, the definition above can be expressed as follows:

\smallskip
\begin{tabular}{l c l c l c l}
$B$ & $\ast$ & $\subseteq$ & $:=$ & $\forall x : S \lamdot (x \in A \Rightarrow x \in B)$ & $:$ & ${\tt prop}$
\end{tabular}

\smallskip
Here ${\tt prop}$ represents the collection of all propositions. Since the symbol `$\subseteq$' is used in the assumption line with identifier $u$, this definition line should precede it. So it could be placed immediately below the assumption lines for $A$ and $B$ given above.

\begin{remark}\label{AutForNots} The newly defined constant \,$\subseteq$ \, depends on three {\it parameters\/}: the context variables $S$, $A$ and $B$; in mathematics, using infix-notation, one would rather write $A \subseteq_S B$ in this definition, instead of mere $\subseteq$.

Yet more common is the notation $A \subseteq B$, so without referring to $S$. It becomes a bit annoying to repeatedly mentioning the `universe' $S$ in the Automath expressions. A mathematician would hardly ever mention these $S$'s -- so leave them implicit. See Section~\ref{IncPar} for how de Bruijn solved this problem.
\end{remark}

In mathematics, definitions are mostly separated from, for example, the statement of a theorem or its proof. In Automath, however, definitions are omnipresent. {\em Every\/} subject $M$ of a typing statement expressed in an Automath line (i.e., a statement of the form $M : N$, expressing `$M$ has category $N$') gets a name, say $d$. In a context named $x$, this is written as\smallskip

\begin{tabular}{l c l c l c l}
$x$ & $\ast$ & $d$ & $:=$ & $M$ & $:$ & $N$
\end{tabular}\smallskip

\noindent (to be read as: `In the context named $x$, constant $d$ is defined as $M$, which has category $N$').

\medskip
Note that the earlier given example -- the definition of $\subseteq$ -- has exactly the same format. We extend this example to a {\it proof\/} of the above LEMMA~5.2\,(a), as expressed in Example~\ref{AutForConExa}. In Automath style this reads as below. We hereby assume that the given lines are preceded by the assumption lines forming the context called $v$, and the definition line for $\subseteq$, given above. We use the derivation rules for the type system $\lambda$D (\citealp{NedGeu}). For example, in the definition line introducing $d_1$ we use that the type of $u$ is $A \subseteq B$, which is definitionally equal to $\forall x : S \lamdot ((x \in A) \Rar (x \in B))$. Then $u\,x\,z$, that is $(u\,x)z$, or $u$  applied to $x$ and the result applied to $z$, is of type (i.e., `proves') $x \in B$.

\medskip
\begin{tabular}{l c l c l c l}
$v$ & $\ast$ & $x$ & $:=$ & $\text{---}$ & $:$ & $S$\\

$x$ & $\ast$ & $z$ & $:=$ & $\text{---}$ & $:$ & $x \in A$\\

$z$ & $\ast$ & $d_1$ & $:=$ & $u\,x\,z$ & $:$ & $x \in B$\\

$z$ & $\ast$ & $d_2$ & $:=$ & $v\,x\,d_1$ & $:$ & $x \in C$\\

$x$ & $\ast$ & $d_3$ & $:=$ & $\lambda z : (x \in A) \lamdot d_2$ & $:$ & $(x \in A) \Rar (x \in C)$\\
$v$ & $\ast$ & $d_4$ & $:=$ & $\lambda x : S \lamdot d_3$ & $:$ & $A \subseteq C$
\end{tabular}

\medskip
The lines defining $d_1$, $d_2$ and $d_3$ can be regarded as the `body of the proof of part(a)' as mentioned in Figure~\ref{AutForFla}.

\medskip
Note the context structure in this example: the context expands from $v$ via $x$ to $z$, and then shrinks again via $x$ to $v$. This can be expressed clearer in the corresponding flag version.

\medskip
We conclude that in an Automath book, apart from assumption lines and primitive lines (see below), {\em every\/} line is a definition line. This is different from what is common in mathematics, but it makes it easy and feasible to write a compact, yet precise formal text. In the above simple example we can$\,$ already$\,$ observe this, since$\,$ we$\,$ use, e.g., $\,d_1\,$ in the$\,$ definition$\,$ of $\,d_2, \,$ and$\,$ so on. Without the
\eject

\noindent
definitions of $d_1$ to $d_3$, the subject for $d_4$ should have been written out as

$\lambda x : S \lamdot \lambda z : (x \in A) \lamdot v\,x\,(u\,x\,z)$,

\noindent which is considerably longer and further away from a proof as is usual in the `mathematical style'.

\smallskip
Another observation is that defined constants such as $\subseteq$ (and also $d_1$ to $d_4$) in the above case, {\em when used\/}, must be provided with a list of {\it instantiated parameters\/}. For example, in the definition above, the type of $d_4$ is $A \subseteq C$. Here $A$ and $C$ are instantiations: in its definition line, $\subseteq$ has the {\em parameter list\/} $S,~A,~B$. But in the final type, the parameter list is $S,~A,~C$. So we have identity substitutions for $S$ and $A$, but a `genuine' substitution of $C$ for the final parameter $B$. This is a simple case; but generally, these substitutions can be quite complicated.

We note that the rules of Automath, being the rules of a type theory, force us to only execute instantiations with expressions of `the same type' as the original context variables. For example, both $B$ and $C$ have type ${\cal P}(S)$. (Often, the instantiation of a context variable requires an analogous substitution of the types; see \citealp{NedGeu}, Section~9.4, for details. For now, we shall not go into this matter.)

\subsection{Primitive lines}\label{AutForPri}

What remains is to explain what lines of sort $(iii)$ stand for. We
call these lines {\it primitive lines\/}. They concern the
introduction of so-called {\it primitive notions\/} (\PN s), used in
stating axioms or primitive elements of sets. For example, the first
two axioms of Peano about the set $\mathbb{N}$ of natural numbers and
a number $0$ in it, can be phrased as follows in Automath-style:

\smallskip
\begin{tabular}{l c l c l c l}
$\epsilon$ & $\ast$ & $\mathbb{N}$ & $:=$ & \PN & $:$ & ${\tt set}$\\
$\epsilon$ & $\ast$ & $0$ & $:=$ &  \PN & $:$ & $\mathbb{N}$
\end{tabular}

\smallskip
Here `{\tt set}' stands for the collection of all sets. So now $\mathbb{N}$ is a primitive set and $0$ a primitive element of type $\mathbb{N}$. For both notions, the context is empty. Note that this notation preserves the typing relation, so these lines express that $\mathbb{N} : {\tt set}$ and $0 : \mathbb{N}$. But in these lines `\PN' is written instead of a definiens after the symbol `:=', since both $\mathbb{N}$ and $0$ are not defined but axiomatically introduced.

\smallskip
One may also define primitive notions {\it in a context\/}. For example, one way of postulating the successor function in Peano-style is as follows (the other way is to introduce $s$ as a function of type $\mathbb{N} \rar \mathbb{N}$, in the empty context):

\smallskip
\begin{tabular}{l c l c l c l}
$\epsilon$ & $\ast$ & $n$ & $:=$ & $\text{---}$ & $:$ & $\mathbb{N}$\\
$n$ & $\ast$ & $s$ & $:=$ & \PN & $:$ & $\mathbb{N}$
\end{tabular}

\smallskip
\noindent and one could continue with defining the number 1 (in the empty context):

\smallskip
\begin{tabular}{l c l c l c l}
$\epsilon$ & $\ast$ & $1$ & $:=$ & $s(0)$ & $:$ & $\mathbb{N}$
\end{tabular}

\smallskip
The last three example lines show all three different sorts of Automath lines: an assumption line, a primitive line and a definition line, respectively.

\medskip
Finally, we note that not only axiomatic {\it objects\/}, but also axioms (i.e., axiomatic {\it propositions\/}) can be expressed in Automath. For example, one of the Peano-axioms can be expressed as follows:

\smallskip
\begin{tabular}{l c l c l c l}
$n$ & $\ast$ & $AX3$ & $:=$ & \PN & $:$ & $\neg(s(n)=0)$
\end{tabular}

\medskip
In an average Automath text, assumption lines and definition lines are the most common ones; they `tell the story'. Primitive lines occur relatively rarely, as one may imagine. For example, the translation (see \citealp{Jutting}) of E.\ Landau's `Grundlagen der Analysis' (\citealp{Lan}) into Automath contains more than a thousand Automath lines, of which only 26 are primitive ones.

\subsection{Special features of Automath}\label{AutForSpe}

The most important characteristics underlying all Automath languages have now been explained. There are four peculiarities in all these languages that must be clarified. We discuss these below.

\subsubsection{$\lambda$-abstraction also for type abstraction}

In most type systems, the Greek letter $\lambda$ is used for {\it functional abstraction\/} and the capital $\Pi$ is used for expressing {\it dependent types\/}. E.g., $\lambda x : A \lamdot M$ expresses the function sending $x$ of type $A$ to $M$, and if $M : N$, then the type of $\lambda x : A \lamdot M$ is $\Pi x : A \lamdot N$. In particular, $\Pi$ can be used for expressing $\forall$ (`universal quantification').

\medskip
Automath, however, does not differentiate between $\lambda$ and $\Pi$. Both $\lambda x : A \lamdot T$ and $\Pi x : A \lamdot T$ are rendered as $[\,x , A\,]\, T$. So the square brackets $[\, \ldots\,]$ have a double role. This is clearly a case of `overloading', that complicates both the meaning and the syntax of the Automath languages.

As to the meaning: the reader of a text must decide for each $[\, \ldots \,]$-abstraction whether it is a usual `functional' one (a `$\lambda$'), or that it is meant to represent a type abstraction (a `$\Pi$'). This is in most cases not very difficult to decide (cf.\ \citealp{Wie}, Section 6.1), but sometimes it asks for special attention.

On the syntactic level, when $\Pi$ is identified with $\lambda$ as in
Automath, we have a difficulty when embedding Automath into the
Calculus of Constructions. We have the choice to map $[\,x , A\,]\, T$
to either $\lambda x : A\lamdot T$ or $\Pi x: A\lamdot T$, and in CC
there are different rules for typing a $\lambda$-abstraction or a
$\Pi$-abstraction. Automath has only one rule for this purpose, but
has the extra notion `type inclusion' (see \citealp{Daa}, Section~4.4)
to mend the ambiguity. Type inclusion allows e.g.\ to move from $\lambda x : A \lamdot s_2$ to $s_2$, and thereby one can
simulate one abstraction rule with the other.

\subsubsection{Incomplete parameter lists}\label{IncPar}

Consider a line of sort $(ii)$ in Automath, so a definition line~$(l)$:

\smallskip
$~~~~x_n~~~\ast~~~c~~~:=~~~M~~~:~~~N$~.

\smallskip
Then both $M$ and $N$ may depend on the variables in the list $x_1, \ldots ,x_n$ referred to by the indicator $x_n$. So, the defined constant $c$ also (but implicitly) depends on its {\em parameter list\/} $x_1, \ldots ,x_n$.

\medskip
Each case this $c$, {\it defined\/} in line~$(l)$, is {\it used\/} in a following line~$(m)$, one must {\it instantiate\/} (i.e., {\it explicitly\/} write out) the expressions $L_1, \ldots ,L_n$ that have to be substituted for the free variables $x_1$ \ldots $x_n$. As we saw before, the result is written as $c(L_1, \ldots ,L_n)$. The list $L_1, \ldots ,L_n$ is called the {\it instantiated parameter list\/} corresponding to that instance of $c$.

\medskip
Now it frequently occurs that such a line~$(m)$ `below' line~$(l)$ has a context that starts with an initial part of the context of $c$, say $x_1 : A_1, \ldots ,x_i : A_i$. Then it regularly happens that $L_1 \equiv x_1, \ldots L_i \equiv x_i$, so the first $i$ substitutions for the context variables are {\it identity substitutions\/}. This often occurs not only in one Automath line, but in a full bundle of lines, and also in other circumstances. In that case it becomes a real burden to mention those initial $x_1, \ldots, x_i$ ever and ever again.

Therefore, Automath has the following {\it parameter omitting\/} convention: an {\it initial\/} part of a parameter list may be omitted when it consists of identity substitutions only. So in the above case, the expression $c(x_1,\ldots,x_i,L_{i+1},\ldots,L_n)$ may be written as $c(L_{i+1},\ldots,L_n)$.

Obviously, the latter parameter list of $c$ is {\it too short\/} to be syntactically correct. But this is easy to repair: if $i$ items are missing in the parameter list, then the first $i$ declared variables from the context should be added in front (in the same order) to make it correct.

\medskip
This convention is very convenient when developing mathematical theories in Automath, since it prevents lots of annoying repetitions of context declarations. So `parameter omitting' is very practical.

\smallskip
In Automath, this convention also applies to {\it primitive constants\/}, having been defined in a primitive line.

\subsubsection{$\beta$-reduction and $\eta$-reduction}\label{BetEta}

As usual in many type systems, also in Automath {\it $\beta$-reduction\/} and {\it $\beta$-convers\-ion\/} are used to cover the notions of function application and convertibility. And again as usual, two terms in Automath that are acceptable and $\beta$-convert to each other, or are `definitionally equal', may be substituted for each other, without explicit justification. (In our example proof in Section~\ref{AutForDef} we showed this, when silently replacing the type of $u$, given as $A \subseteq B$, by its definiens $\forall x : S \lamdot ((x \in A) \Rar (x \in B))$, in order to obtain $uxz$ as inhabitant of $x \in B$.)

\medskip
We recall that $\beta$-reduction is the compatible closure of the relation expressing `applying a function to an argument', i.e., $(\lambda x : M)N \rar_\beta M[x := N]$, where $M[x := N]$ is the result of substituting $N$ for all free $x$'s in $M$. Conversion is the equivalence relation generated by $\beta$-reduction. Since $\lambda$s and $\Pi$s are identified in Automath, $\beta$-reduction and $\beta$-conversion also apply to $\Pi$-types. Such $\Pi$-types may be provided with arguments in Automath, which is very uncommon in other type theories, but a consequence of the $\lambda$-$\Pi$-identification.

\medskip
Automath not only has $\beta$-reduction, but also allows {\it $\eta$-reduction\/} of terms. This corresponds to the `extensionality' relation on $\lambda$-terms, generated by the reduction $(\lambda x : A \lamdot N x) \rar_{\eta} N$ (if $x$ is not a free variable in $N$). We note that in Jutting's Automath-version of Landau's book \citep{Lan}, only two $\eta$-reductions are necessary, whereas over 6.000 $\beta$-reduction take place in the checking procedure and more than 20.000 definition unfoldings ({\it `$\delta$-reductions'\/}). Again, see \citealp{Jutting}.

\subsubsection{Argument precedes function}

De Bruijn advocates inversion of the usual order function--argument in a formal treatment of mathematics. Instead of $f a$ for `$f$ applied to $a$\,' (the usual notation in $\lambda$-calculus), he prefers $\!<\! a \!>\!f$ and uses this in the Automath languages. He motivates this as follows: an abstraction is written in front ($M$ becomes $(\lambda x : A) \lamdot M$ when abstracted over $A$), so it is consistent to do the same with an application ($M$ becomes $\!<\!A\!>\!M$ when applied to $A$).

This notation for application has both technical and philosophical advantages, which we shall not mention here. See e.g.\ \citealp{KamNed}. But it is unusual and may be confusing to work with.

\subsection{Interpretation of the example text}\label{AutForInt}

Now it is easy to `read' Jutting's Automath text as reproduced in Figure~\ref{AutForLan}. Below we give its translation in common mathematical language. As in Automath, we use round brackets for instantiation and pointed brackets for application. Some explicit notes are added between square brackets. Note that the translation doesn't give many clues about the variations in the contexts of the various lines, whereas the Automath text is precise about these matters.

{\small
\begin{quotation}

(1) Let $A$ be a proposition

(2) and let also $B$ be a proposition.

(3) Then $\IMP$ is defined as the function space $\lambda x : A \lamdot B$. [This $\lambda$ must be read as $\Pi$, so $\IMP$ (in the context of $A$ and $B$) can be considered to embody the implication $A \Rar B$.]

(4) Let moreover $C$ be a proposition,

(5) assume that $\IMP(A,B)$ holds with proof object $I$

(6) and assume that $\IMP(B,C)$ holds with proof object $J$.

(7) Then $\lambda X : A \,\lamdot\, ((J\,I)X)$ proves $\IMP(A,C)$ and is called $\TRIMP$. [$\lambda$ represents $\Pi$, again. See also Section~\ref{AutForSpe},\,(4).]

(8) By axiom, $\CON$ is a proposition, in the empty context. [It formalizes the notion of `contradiction'.]

(9) In the context of $A$, constant $\NOT$ is defined as $\IMP(\CON)$. [To be read as $\IMP(A, \CON)$, since the given parameter list is incomplete.]

(10) In the same context, $\WEL$ is defined as $\NOT(\NOT(A))$.

(11) Assume that proposition $A$ is inhabited by $A1$,

(12) then $\lambda x : \NOT(A) \lamdot (X \, A1)$, called $\WELI$, proves $\WEL(A)$ [so lines (11)and (12) together express: if $A$ holds, then also $\neg \neg A$].

(13) Assume that $\WEL(A)$ is inhabited by $W$,

(14) then, by axiom, $A$ is inhabited by $\ET$ [`excluded third'].

(15) Assume that $\CON$ is inhabited by $C1$,

(16) then $\ET(\lambda x : \NOT(A) \lamdot C1)$ proves $A$. [Note: the parameter list of $\ET$ is incomplete.]
\end{quotation}

}

\section{Automath and modern type theory}\label{AutMod}

In this section we compare the essence of the syntax of Automath (which we called the `generic' syntax) to that of another type system: the {\it Calculus of Constructions\/} of Coquand and Huet ({\it CC\/}, also called $\lambda$C). Cf.\ \cite{CoqHue}. This CC is a very influential type theory, both for theoretical purposes as in practice. It is, for example, the basis of the theorem prover Coq (see \citealp{BerCas}). CC has been devised around 1984, many years later than Automath (dating from 1968). However, Coq is still in development (see \citealp{Coq}), but the Automath work has come to an end. The standard reference on type systems is \cite{Bar1992}.

\medskip
Coquand and Huet pay tribute to de Bruijn's Automath \citep{CoqHue85}, but the original syntactic rules of Automath are rather complicated and difficult to compare with CC's elegant syntactic rules. An early example in this direction can be found in \citealp{Daa}, 5.5.4. An extension of CC with definitions, such as the type system $\lambda$D described in \cite{NedGeu}, has more direct links to Automath. But the direct relation of $\lambda$D to Automath that we show below, has not yet been described.

\subsection{Abandoning inessential Automath features}\label{AutModAba}

First we recall that the purpose of N.G.\ de Bruijn when developing Automath, was to produce a {\it practical\/} system, immediately useable for a mathematician to formalize and verify mathematical content. A magnificent example to show that de Bruijn attains his goals, is the impressive translation of the complete book {\em Grundlagen der Analysis\/} (`Foundations of Analysis') of E.\ Landau \citealp{Lan}, into Automath, by L.S.\ van Benthem Jutting. The translation of the book is well readable after a little exercise, although it covers 13.433 Automath lines. (In the previous section we gave a very short impression of this translation.)

\medskip
For the conversion of the Automath syntax to a CC-like version, human-readability is less important then syntactical simplicity. That's why we {\it abandon\/} some complicating features of Automath (cf.\ Section~\ref{AutForSpe}) that are a bit unusual, or only meant for the human user, namely:

\medskip

(1) {\it $\lambda$-abstraction also for type abstraction\/}~  This Automath-peculiarity is not very attractive and unnecessary (\citealp{Wie}). We restore $\Pi$-abstraction in our CC-like derivation rules.

\smallskip
(2) {\it Incomplete parameter lists\/}~  These are desirable for humans, but needless for theoretical analysis. So we abandon them.

\smallskip
(3) {\it $\eta$-reduction\/}~  This can be harmlessly omitted, as suggested already in Section~\ref{BetEta}.

\smallskip
(4) {\it Argument precedes function}~  This is an Automath-peculiarity that is debatable because it is so uncommon. We restore the usual order. We also employ the usual $\lambda$-calculus notation. So we write {\em application\/} of $F$ to $M$ as $F \, M$ (with an invisible operation sign between function and argument), instead of $\!<\!M\!>\!F$ in Automath. Note that we also use the notation with round brackets, $G(N)$, but only for {\em instantiation\/}. So $G(N)$ has the meaning: `constant $G$ with $N$ substituted for the context variable'.

\subsection{Automath judgements}\label{AutModJud}

We have discussed the three kinds of Automath lines: {\it assumption lines, definition lines and primitive lines\/}. Now we want to give a type system for establishing {\it correctness\/}. Note that, in an Automath text, the lines are not given as a set, but as an ordered {\it list\/}. Such a list is called a {\it book\/} in the original Automath report. An Automath book is {\it correct\/} if it can be derived by means of the rules of Automath.

\medskip

There are different correctness requirements for the three sorts of lines in Automath. First note that the lines in a correct Automath book depend on earlier ones. We mention two examples: (1) an existing context can be extended or reopened later by referring to its context indicator; (2) a defined constant can be used in every line following its definition, providing that it has been properly provided with instantiations for its parameters.

These dependencies are enabled by the ordered character of each Automath book: every line may depend on any number of {\it previous\/} lines. If lines~(1) to $(l)$ form a correct (and ordered) book ${\bf B}$, then we may add a new line~$(l + 1)$, provided that it is correct with respect to ${\bf B}$. So books are developed {\em incrementally\/}, line by line. (This linear build-up is not in contradiction with the tree-like build-up in common type systems, as we sufficiently explained in \citealp{NedGeu}.)

\smallskip
We express correctness not as in the original Automath report (\citealp{deB1970}), but by means of three specific rules. We refer to the system $\lambda$D, as described in \citealp{NedGeu}, to obtain full correctness of an Automath text.

The system $\lambda$D is a natural extension with definitions of $\lambda$C (which is an alternative name for CC). We recall the following important notion used in type systems as $\lambda$C (cf.\ \citealp{Bar1992} or \citealp{BarGeu}).

\medskip
A type-theoretical {\em judgement\/} has the form

\smallskip
$\Gamma ~\vdash~ M \, : \, N$,

\smallskip
\noindent expressing that it is derivable (`\,$\vdash$\,') in context $\Gamma$ that $M$ has type $N$. Here, a context is a list of variable declarations $x_1 : A_1, \ldots, x_n : A_n$.

\medskip
Since we have to do with an Automath book, we also have to consider a context $\Delta$ of definitions. So in order to establish correctness, we need for Automath some sort of {\em extended judgement\/}. This has in $\lambda$D the following form:

\smallskip
$\Delta ~;~\Gamma ~\vdash~ M \, : \, N$,

\smallskip

\noindent meaning that, with respect to definition context $\Delta$ and context $\Gamma$, we can derive that $M$ has type $N$. Here, a definition context is a list of definitions, each of the form $c(x_1 : A_1, \ldots, x_n : A_n) := M : A$. (A lemma about the deduction rules implies that $\Delta$ and $\Gamma$ have already been justified in that case; see again \citealp{NedGeu}.)

\subsection{Coherent Automath books}\label{AutModCoh}

We start with a formal description of an Automath {\em book\/}, referring back to notions introduced in Section~\ref{AutFor}.

The following notions are borrowed from type theory (see e.g.\ \citealp{Bar1992}): {\it Var\/} is an infinite set of {\it variables\/}; {\it Const\/} is a disjoint, also infinite set of {\it constants\/};
{\it Term} is the set of {\it pseudo-terms\/} of $\lambda$D, given as usual by an inductive definition.

{\it Term} includes {\it Var\/}, {\it Const}({\it Term}, \ldots, {\it Term}), $\lambda {\it Var} : {\it Term} \lamdot {\it Term}$ (the so-called $\lambda$-terms), $\Pi {\it Var} : {\it Term} \lamdot {\it Term}$ (the $\Pi$-terms) and ({\it Term} {\it Term}) (the application terms).

\begin{definition}[Line; book]
\begin{enumerate}
\item An {\em Automath line\/} is either an {\em assumption line\/}, a
  {\em definition line\/} or a {\em primitive line\/}. See
  Figure~\ref{threespecies} for the appearance of each of these lines
  and for the various notions connected with Automath lines,
  e.g.:\\ An {\it indicator\/} is $\epsilon$ or an $x \in {\it Var}$,
  an {\it identifier\/} is an $x \in {\it Var}$ or a $c \in {\it
    Const}$, a {\it content\/} is either `$\--$' or `\PN' or an $M \in
  {\it Term}$, and a {\it category\/} (or {\it type\/}) is an $A \in
  {\it Term}$.
\item A sequence of Automath lines is called a {\em book\/}, denoted
  ${\bf B}$.
\end{enumerate}
\end{definition}

We give some terminology around lines and books:

\begin{definition}[$\in$, $+$, $<$, $\subseteq$; precede, subbook]
\begin{enumerate}
\item $l \in {\bf B}$ means that line $l$ occurs in book ${\bf B}$.
\item Concatenation of lines and/or books is denoted by the symbol
  $+$, e.g., ${\bf B_1} + {\bf B_2}$ or ${\bf B_1} + l$.
\item We say that line $l$ {\em precedes\/} line $m$ in book ${\bf B}$,
  denoted $l < m$, if ${\bf B} \equiv {\bf B_1} + l + {\bf B_2} + m +
  {\bf B_3}$.
  \eject
\item We say ${\bf B'}$ is a {\it subbook\/} of ${\bf B}$, and write
  ${\bf B'} \subseteq {\bf B}$, if all lines in ${\bf B'}$ also appear
  in ${\bf B}$ and inherit the precedence order.
\end{enumerate}
\end{definition}

A first step to correctness concerns the structure of Automath texts as regards the use of {\em identifiers\/} introduced in an Automath book. First, we prevent ambiguities in referencing, which occur when an identifier is introduced multiply. Therefore we require that each identifier is {\em unique\/}.
(This restriction can be eliminated, e.g.\ by a paragraph system; cf.\ Section \ref{AutForAss}.)

We also must ensure that no indicator occurring in an Automath line does `dangle': we recall that each context indicator $z$ must refer to precisely one `earlier' introduced identifier of the same name, in order to establish exactly which context is `named' by $z$. This earlier identifier should pop up in an assumption line.

An Automath book in which the last-mentioned conditions hold, we call a {\em coherent\/} book.

\begin{definition}[Coherent book; $\prec$)]\label{coherent}
\begin{enumerate}
\item The book ${\bf B}$ is called {\em coherent} if the following hold.
\begin{enumerate}
\item All identifiers of lines in ${\bf B}$ are different.
\item Each context indicator $y$ in a line $m \in {\bf B}$, with $y
  \not \equiv \epsilon$, appears as {\em identifier\/} in an {\em
    assumption\/} line $l$ of ${\bf B}$ where $l < m$.
\end{enumerate}
\item Let $m$ be a line in a book ${\bf B}$, with indicator $y$ and
  identifier $z$. If ${\bf B}$ is coherent and $y \not\equiv
  \epsilon$, then we say that $y \prec z$.
\end{enumerate}
\end{definition}

We have the following easy properties. If $m$ is a line in a coherent book ${\bf B}$ with identifier $z$ and indicator $y \not \equiv \epsilon$, then there is exactly one line $l < m$ in ${\bf B}$ with identifier $y$. So, for each identifier $z$ in a coherent book ${\bf B}$, we can `follow the trail back' to collect the full chain of variables that `precede' $z$ according to the relation $\prec$. Note that such a chain is always finite and consists of variables, but for the `greatest' element, which may be a constant.

\medskip
If such a $z$ {\em is\/} a variable (not a constant), we recursively define the notion $\gamma_z$ as the full list of the variables in this chain, including $z$ itself. We simultaneously define the typing-context $\Gamma_z$ of $z$, being $\gamma_z$ provided with typing information for each of the variables in it.

\begin{definition} [Subject-list $\gamma_z$; typing-context $\Gamma_z$]\label{subbook}

Let ${\bf B}$ be a coherent book and let {\em assumption\/} line $m$ be in ${\bf B}$. We proceed by induction on the relation `$\prec$', taking it that $\gamma_\epsilon$ is the empty list and $\Gamma_{\epsilon}$ the empty context.
Let \mbox{$m \, \equiv \, y \,  \ast \, z \, := \, \text{---} \, : \, A$}, where $y$ may be $\epsilon$. If $y \not\equiv \epsilon$, then $y \prec z$, so $\gamma_y$ and $\Gamma_y$ are already known by induction. Now we define:

\medskip
(1) The {\em subject-list\/} $\gamma_z$ of $z$ is \,$\gamma_y, \, z$.\smallskip

(2) The {\em typing-context\/} $\Gamma_z$ of $z$ is  \,$\Gamma_y, z \!:\! A$.
\end{definition}

\subsection{The derivation rules for Automath}\label{AutModThr}

We will now define what a {\em well-formed book\/} is in the Automath sense, using the system $\lambda$D to obtain the desired type-correctness for the terms. As $\lambda$D contains rules for definitions, it fits very well with Automath, but basically any other type theory with formal definitions could be used. This is because the book structure with lines, built up in the way described in the previous section, is largely orthogonal to the notion of {\em well-typedness\/}, that we need to judge whether a new line can be added to a book.

\medskip
We define by induction on ${\bf B}$ the notion ${\bf B} ~ {\it ok}$,
expressing that ${\bf B}$ is well-formed. The definition splits into
three parts, depending on whether we add {\em (i)} an assumption line,
{\em (ii)} a definition line or {\em (iii)} a primitive line to a
book~${\bf B}$, which we assume to be already {\em ok\/}. The new book
that we obtain after the addition of one of the three lines, we call
${\bf B}^+$. Simultaneously we define $\Delta_{\bf B}$, the
translation of the well-formed book ${\bf B}$ into a $\lambda$D-`book'
(see \citealp{NedGeu}).

We recall that the symbol $s$ stands for a so-called {\em sort\/}, being either $\ast$ or $\Box$. These things have been described and explained in \cite{Bar1992}. See also, again, \citealp{NedGeu}.

\begin{definition}[{\em ok\/}; translation $\Delta_{\bf B}$]\label{adding}
\begin{enumerate}
\item {\em Start of the induction:}
\begin{itemize}
  \item the empty book $\epsilon$ is ok
  \item $\Delta_\epsilon \, := \,  < ~ > $
\end{itemize}
\item
\begin{itemize}
\item[$(i)$] If ${\bf B}^+ = {\bf B} +  (z \, \ast \, x \, := \, ~  - \hspace{-1.2em} - \hspace{-1.2em} - \, \, : \, A)$, i.e., {\em adding an assumption line to ${\bf B}$}, then
\begin{itemize}
\item
  $(\mbox{\em assum}) \quad
  \begin{prooftree}
    {\bf B} ~{\it ok} \qquad \Delta_{\bf B}\,;\,\Gamma_z \, \vdash\, A : s
    \justifies
        {{\bf B}^+} ~ {\it ok}
        \using{\mbox{if }x\mbox{ is ${\bf B}$-fresh}}
  \end{prooftree}$
  \item $\Delta_{{\bf B}^+} \, := \, \Delta_{\bf B}$
\end{itemize}

\item[$(ii)$] If ${\bf B}^+ = {\bf B} \hspace{0.02em} + \hspace{0.02em} (z ~ \ast ~ c ~ := \, M \, : \, A)$, i.e., {\em adding a definition line to ${\bf B}$}, then
\begin{itemize}
\item
  $(\mbox{\em defin}) \quad
  \begin{prooftree}
    {\bf B} ~{\it ok} \qquad \Delta_{\bf B}\,;\,\Gamma_z \, \vdash\, M : A
    \justifies
        {{\bf B}^+} ~ {\it ok}
        \using{\mbox{if }c\mbox{ is {\bf B}-fresh}}
  \end{prooftree}$
\item $\Delta_{{\bf B}^+}  \,  := \, \Delta_{\bf B} \, ,  (\Gamma_z \rhd c(\gamma_z) \, := \, M  :  A)$
\end{itemize}

\item[$(iii)$] If ${\bf B}^+ = {\bf B} \hspace{0.02em} + \hspace{0.02em} (z ~ \ast ~ c \, := \,  \PN \, : \, A)$, i.e., {\em adding a primitive line to ${\bf B}$}, then
\begin{itemize}
\item
  $(\mbox{\em prim}) \quad
  \begin{prooftree}
{\bf B} ~{\it ok} \qquad \Delta_{\bf B}\,;\,\Gamma_z \, \vdash\, A : s
    \justifies
        {\bf B}^+ ~ {\it ok}
        \using{\mbox{if }c\mbox{ is ${\bf B}$-fresh}}
  \end{prooftree}$
\item $\Delta_{{\bf B}^+} \, := \, \Delta_{\bf B} \, ,  (\Gamma_z \rhd c(\gamma_z) \, := \, \PN \, : \, A)$
\end{itemize}
\end{itemize}
\end{enumerate}
\end{definition}

The use of $\lambda$D as an `auxiliary type system' to guarantee that the lines added to the book contain well-formed (and well-typed) expressions, conforms with the philosophy of de Bruijn: he viewed the type checking of the expressions that were added to a book as a purely algorithmic issue. In his 1970 paper, de Bruijn introduces the book-concept and describes in an algorithmic way what should be verified (i.e., type-checked) before a line can be added. Our Definition~\ref{adding} mimics this via $\lambda$D-derivation rules.

\subsection{Automath and type theory}\label{AutTypThr}

In type theory, assumptions (or `declarations', in type-theoretic words) are usually collected into `contexts' $\Gamma \, \equiv x_1 : A_1, \hdots , x_n : A_n$, preceeding a `judgement'. But in an Automath book ${\bf B}$, every new assumption is written in a separate line. Therefore, it may easily happen that an assumption turns out to be {\em redundant\/}, to be explained below. We firstly define:

\begin{definition}[dead-end assumption line]\label{RedAss}

An assumption line  $l ~ \equiv ~ z \, \ast \, x \, := \, \text{---} \, : A$ in a book ${\bf B}$ is a {\em dead-end\/} when identifier $x$ is not an indicator of any line following $l$ in ${\bf B}$.
\end{definition}

One might say that a dead-end assumption line has no `successors' in the book. Of course, there may be several dead-end lines in a book ${\bf B}$. It will be clear that each dead-end line can be removed from the book without influencing any of the other lines. But note that, after removal of a dead-end-line, another dead-end line can turn up that first was `hidden.

\smallskip

\begin{example}\label{ExaDea}

Assume that lines $l_1$ and $l_2$ appear in ${\bf B}$, with $l_1 < l_2$ and such that

\vspace{0.3em}

$l_1 ~\equiv ~ z \, \ast \, y \, := \, \text{---} \, : A$ ~ and ~
$l_2 ~\equiv ~ y \, \ast \, x \, := \, \text{---} \, : B$.

\smallskip

\noindent Now, let $l_2$ be a dead-end assumption line in ${\bf B}$ and let's remove this $l_2$. Then line $l_1$ may become a new dead-end in ${\bf B}$ (which is was not before). This happens if identifier $y$ does not occur as indicator in another line than the now removed $l_2$.
\end{example}

It is not hard to show that one may successively remove {\em all\/} dead-ends from a given book ${\bf B}$, also the newly popped-up as described just now. All these lines together are called the {\em redundant\/} assumption lines of ${\bf B}$. It also holds that there is a {\em unique\/} final result, independent of the {\em order\/} of the removals of dead-ends. (Of course, `new' dead-ends such as the above $l_2$ can only be removed neatly {\it after\/} the removal of the `old' ones such as $l_1$.) Moreover, in each step of this process of removals, one coherent book is replaced by another coherent one, and an {\it ok\/}-book by a new {\it ok\/}-book.

Given a coherent book ${\bf B}$, we call the result of eliminating all dead-ends from it a {\em clean book\/}.

\begin{definition}[clean book; {\tt cl}(${\bf B}$)]\label{CleBoo}
\begin{enumerate}
\item A coherent Automath book ${\bf B}$ in which no assumption line
  is a dead-end, is called a {\em clean\/} book.
\item The clean book being the result of eliminating all redundant
  lines from a book~${\bf B}$, is denoted ${\tt cl}({\bf B})$.
\end{enumerate}
\end{definition}

The following theorem links an Automath book ${\bf B}$ to the type-theoretic book $\Delta_{\bf B}$. It follows directly from the rules and the definitions in Definition~\ref{adding}. The correspondence between the two kinds of books is obvious and doesn't need further discussion.

\begin{theorem}\label{simlem}
Let ${\bf B}$ be an {\em ok\/} Automath book.
\begin{enumerate}
\item Then $\Delta_{\bf B}$ is a `legal environment' in $\lambda$D, in
  the sense as defined in \cite{NedGeu}, Def.~10.4.
\item Moreover, the definition lines of ${\bf B}$ correspond
  one-to-one with the judgements in $\Delta_{\bf B}$, in the sense
  that the sequence of constants in the definition lines of ${\bf B}$
  is the same as the sequence of constants in the judgements occurring
  in $\Delta_{\bf B}$. Also, the sequence of typing expressions $M :
  A$ in the definition lines ${\bf B}$ is the same as in the
  judgements of $\Delta_{\bf B}$, and a similar thing holds for the
  typing property $\PN : A$.
\item If ${\bf B}$ is a {\em clean\/} book, then all assumption lines
  with identifier $x$ of ${\bf B}$ occur in the contexts of exactly
  those judgements for which the constant $c$ depends on $x$ (i.e., $x
  \in \gamma_y$, where $y$ is the indicator of the $c$-line in ${\bf
    B}$.
\end{enumerate}
\end{theorem}

So there is a strong connection between a clean Automath-book ${\bf B}$ and the corresponding type-theoretic book $\Delta_{\bf B}$. It is not hard to see that all relevant `information' in ${\bf B}$ (i.e., the information in ${\tt cl}({\bf B})$) has a counterpart in $\Delta_{\bf B}$. Note that the image $\Delta_{\bf B}$, being an environment in $\lambda$D, consists of definitions only.

In the transition from ${\tt cl}({\bf B})$ to $\Delta_{\bf B}$, the nice `sharing' of context declarations $x : A$, an asset of Automath, is completely undone. Every definition in $\Delta_{\bf B}$ has its own context $\Gamma$, leading to numerous repetitions of declarations and no sharing at all.

\medskip
The lack of sharing is also obvious in the simplest translation the other way around, leading from a legal environment $\Delta$ to an Automath book ${\bf B}_\Delta$. In this translation, the image of a definition in $\Delta$, e.g., $\overline{x} \, : \, \overline{A} \, \triangleright \, a(\overline{x}) \, := \, M \slash \PN \, : \, N$, becomes a list of $n$ assumption lines ending in one definition line or primitive line (here $n$ is the number of declarations in $\overline{x} \, : \, \overline{A}$):

\begin{figure}[!b]\small
\centering
\begin{tabular}{|l r  c  l  c  l  l  l|} \hline
(1) & & $\ast$ & $A$ & $ := $ & --- & ; & {\em Prop}\\
\multicolumn{8}{|l|}{ $\longrightarrow$ {\em Check:\/} $ \emptyset$  ;  $\emptyset$  \,$\vdash$\,  {\em Prop}  :  {\em Type}}\\ \hline
(2) & $A$ & $\ast$ & $B$ & $ := $ & --- & ; & {\em Prop}\\
\multicolumn{8}{|l|}{$\longrightarrow$  {\em Check:\/} $ \emptyset$  ;  $A : $ {\em Prop}  \,$\vdash$\,  {\em Prop}  :  {\em Type}}\\ \hline
(3) & $B$ & $\ast$ & $\IMP$ & $ := $ & $\Pi x : A \lamdot B$ & ; & {\em Prop}\\
\multicolumn{8}{|l|}{$\longrightarrow$  {\em Check:\/} $ \emptyset$  ;  $A,\, B : $ {\em Prop}  \,$\vdash$\,  $\Pi x : A \lamdot B$ : {\em Prop}}\\ \hline
(4) & $B$ & $\ast$ & $C$ & $ := $ & --- & ; & {\em Prop}\\
\multicolumn{8}{|l|}{$\longrightarrow$  {\em Check:\/} $\IMP(A,B)  :=  \Pi x : A \lamdot B : $ {\em Prop}  \,;\,}\\
  \multicolumn{8}{|l|}{\hspace{2.4cm} $A,\, B : $ {\em Prop}  \,$\vdash$\,  {\em Prop} : {\em Type}}\\ \hline
(5) & $C$ & $\ast$ & $I$ & $ := $ & --- & ; & $\IMP(A,B)$\\
\multicolumn{8}{|l|}{$\longrightarrow$  {\em Check:\/} $\IMP(A,B)  :=  \Pi x : A \lamdot B : $ {\em Prop}  \,;\,}\\
 \multicolumn{8}{|l|}{\hspace{2.4cm}$A,\, B,\, C : $ {\em Prop} \,$\vdash$\,  $\IMP(A,B)$ : {\em Prop}}\\ \hline
\end{tabular}
\caption{Automath text with checking obligations}
\label{AutForLanChe}
\end{figure}

\begin{table}[h]
\centering
\begin{tabular}{l c l c l c l}
$\epsilon$ & $\ast$ & $x_1$ & $:=$ & \text{---} & $:$ & $A_1$\\
$x_1$ & $\ast$ & $x_2$ & $:=$ & \text{---} & $:$ & $A_2$\\
$\vdots$ &&&&&& \\
$x_{n-1}$ & $\ast$ & $x_n$ & $:=$ & $\text{---}$ & $:$ & $A_n$\\
$x_n$ & $\ast$ & $a$ & $:=$ & $M \slash \PN$ & $:$ & $N$
\end{tabular}
\end{table}

In the desired translation from $\Delta$ to ${\bf B}_\Delta$, one replaces the lines in $\Delta$, one by one and in the same order, by clusters of lines as given just now. Of course, in this process one probably has to rename variables and constants in order to keep all variables and constants distinct.

\begin{lemma}\label{LemlDtoAut}
The translation described above maps a legal $\lambda$D-environment $\Delta$ onto an {\em ok\/} and clean Automath book ${\bf B}_\Delta$.
\end{lemma}

It is not hard to find a procedure to recover the sharing of variables after the translation. We don't elaborate on this.

\subsection{Example of the verification of an Automath book via a typed lambda calculus}

We show how the first five lines of the Automath text of Figure~\ref{AutForLan} from van Benthem Jutting's formalisation of Landau's book, is a `correct' book in the sense of Definition~\ref{adding}. See Figure~\ref{AutForLanChe}.

In this example, we write the text line by line and indicate in between the $\lambda$D-typing judgement that is checked. We omit the `{\em ok\/}'-word, but just incrementally add lines. We write, as nowadays usual, $\Pi x : A \lamdot B$ instead of $[x : A]B$.
Note for example, that line~(5) may be added to the book consisting of lines (1)---(4), because the judgement following line~(5) in Figure~\ref{AutForLanChe} -- the one to be checked -- can be derived in type theory (e.g., $\lambda$D). (This judgement does not belong to the Automath text, it is added just for the example.) For the exact derivation rules concerning the derivation in $\lambda$D, see again \citealp{NedGeu}.

\section{Conclusion}\label{Conc}

\subsection{De Bruijn's philosophy concerning Automath}\label{AutModPhi}

As we have discussed already in the first part of this paper, N.G.\ de Bruijn only allows three species of lines in an Automath book. In our terminology in Section~\ref{AutMod}, these
are recorded in the conclusions of the rules {\em (assum)}, {\em (defin)} and {\em (prim)}: each of these statements enables the extension of
the {\em ok\/}-book ${\bf B}$ with one of the three kinds of Automath lines to generate a new {\em ok\/}-book, ${\bf B}^+$.

One might wonder where the justifications of the judgements in the
{\em premisses\/} of these rules are registered, the $\lambda$D-judgements in
Definition~\ref{adding}. These judgements are not part of an Automath book. To
explain this, we should understand de Bruijn’s original ideas, since
he had a number of strong opinions about the formalization of
mathematics and the desirability of such a formalization.

\smallskip
(1) Right from the actual start of the computer age, in the 1960's, N.G.\ de Bruijn recognised {\em the potential of computer use for different kinds of mathematical work\/} -- and in particular for verification of mathematical content. That ignited his drive to develop a formal language for mathematics, since he realised that computer aid could considerably enhance the credibility of verification. His early attempts in this direction resulted in the development of a formal language for mathematics, which he called Automath (`mathematics aided by an automaton').

\smallskip
(2) Being a genuine (and very respected) mathematician, it is needless to say that de Bruijn produced a meticulous description of the syntax of Automath. However, {\em he did not shun computer assistance\/} in the actual verification of Automath syntax.

For that purpose, he asked computer scientist I.\ Zandleven to write a large computer programme, with which indeed Jutting's voluminous Automath translation of Landau's analysis textbook (\citealp{Lan}) has been checked. In essence, this Automath programme verified the derivability of the checking requirements. Indeed, de Bruijn's opinion was that such {\em `administrative' work could well be left to a computer\/}. And the more so, since a bit later it was proved that the Automath syntax could be seen as a {\em decidable\/} derivation system.

\smallskip
(3) In de Bruijn's philosophy, an Automath book should be formal and formally verifiable, but {\em it should follow the original mathematical reasoning as closely as possible\/}. At the same time, Automath should be {\em as simple as possible\/}. Therefore he endeavoured to try to limit himself, and he attempted to do the job with assumptions and definitions only. He realised, of course, that mathematics offers more than these two `linguistic categories'. For example, the majority of mathematical texts consists of {\em statements\/}, in order to bring the reasoning a step further. De Bruijn's invention was to {\em extend\/} all usual mathematical statements to a definition (cf.\ what we said about this in Section~\ref{AutForDef}) and to also treat axioms or axiomatic notions as a kind of definitions (`primitive notions').

\smallskip
(4) Thus he could achieve his goal to {\em limit himself to only assumption lines, definition lines and primitive lines\/} in the formal Automath texts, and yet stay close to the mathematical written discourse in its usual layout and order. On a small scale, this is easily visible in our mathematical-style phrasing of the tiny part of Jutting's original Automath text, pictured in Figure~\ref{AutForLan}, when comparing this to the `mathematical' text in Section~\ref{AutForInt}. Looking back to the central part of the present paper, we might say that a {\em human\/} should compose the Automath text and that the {\em computer programme\/} mainly works on checking the {\em type-theoretic premisses\/} of the three main derivation lines (see Section~\ref{AutModThr}), inspecting the Automath lines one by one.

\smallskip
(5) A remark in this respect is, that despite the decidability, the computer program used in the seventies sometimes had considerable problems with its verification task, in particular because of the (occasionally almost forbidding) size of the `search space' involved in the definitional equality checking procedure. In such cases, de Bruijn had no problem with an intervention, amounting to an auxiliary lemma (plus proof!) or an extra definition, in order to {\em bridge the verifiability gap between one Automath line and another one\/}. All this fitted smoothly in the Automath syntax, so he saw no objection in that little help. Such a small intervention was incidentally executed, for example, in the Landau-translation by L.S.\ Jutting.

\subsection{Other proof checkers}

At the time when de Bruijn developed Automath, the idea of computer
assisted formal proof checking arose at various other places. There
has been some contact between these developments, but mostly these
were done independently. Only later, in approximately the 1980s, there
was more international exchange on the ideas and concepts underlying
these projects, and the developers sometimes used ideas from each other
or developed new systems by combining features from various existing
ones.

We will not provide an overview of proof assistants here, but
give a brief overview of related projects and how they compare
with the `book' approach that de Bruijn took with Automath. For a
historical overview of proof assistants and a more in depth discussion
of the ideas, see \citealp{Geu2009}; for a concrete picture of what
present day proof assistants are and how one interacts with them, see
\citealp{Wie2006}, where 17 provers are compared by showing the
formalization of the basic mathematical result that $\sqrt{2}$ is
irrational.

\subsubsection{Mizar}

The Mizar system (see \citealp{Miz}) has been developed by
Andrzej Trybulec and his co-workers since around 1970, with as aim to
create a library of formalised mathematics. The user enters a
mathematical text (definitions, statements, proofs) and the system
checks that text by verifying the logical steps made in the proof and by
checking the definitions for syntactic correctness.

The idea is close
to de Bruijn's idea of a {\em mathematical vernacular\/}, where the mathematics
that is entered into the system should be close to what one would
write in a book or an article. The idea that the checking is going on `behind
the scene', while the user feeds new mathematics line-by-line, is
similar to the book approach of Automath, but in Mizar this is usually done in a
`batch' process: a user writes an article and feeds that as a
whole to Mizar, and then it is the system that reports which logical steps it
can't follow. Mizar has existed continuously since the 1970s and has
developed a very large library of formalised mathematics, {\em MML\/}.

\subsubsection{HOL}\label{Hol}

The HOL proof assistants (see \citealp{GorMel}) are
based on the LCF approach (\citealp{Gor}) due to R.\ Milner in the
beginning of the 1970s. The idea is to have an abstract data type of
theorems, where the only closed terms of this data type correspond to
provable formulas in higher order logic. (Basically, this logic is the higher order
predicate logic as defined by A.\ Church, (\citealp{Chu}), based on his
simple theory of types.) The functions one can write for producing
terms of type `theorem' correspond to the derivation rules for high
order logic, which guarantees the validity.

The approach is quite
different from de Bruijn's, as one can write any function (in the
programming language that the type `theorem' resides in, being ML in the
original approach of Milner) to produce terms of type `theorem'. These
are basically proof-tactics, so one does not write proofs in
mathematical style, but in programming style. The most used systems in
the `HOL family' are HOL4, HOL-light and ProofPower.

\subsubsection{Isabelle}

The Isabelle proof assistant
(\citealp{Nip}), developed in Cambridge (Paulson) and Munich
(Nipkow, Wenzel), is also based on the LCF approach and it caters for
various logics to be definable in it. The main used variant is
Isabelle/HOL, which is the higher order logic mentioned in \ref{Hol}. A
particular feature of Isabelle is the so called `Isar mode', which is
an input mode for Isabelle similar to Mizar, where one enters
mathematical texts in vernacular style, which is then processed by the
system: definitions are checked for correctness and stored, and the
proof steps are verified. This is again similar to the `book'
approach, with various advanced features giving additional proof
automation support.

\subsubsection{Coq}

In the beginning of Section~\ref{AutMod}, we already mentioned the Coq proof assistant
(\citealp{BerCas}; \citealp{Coq}), which is also based on dependent type theory
and the propositions-as-types principle, like Automath. Also in the
Coq system, the type checking algorithm is behind the scenes, while
the user writes mathematics. The definitions are entered and type
checked, and stored as definitions, much like in $\lambda$D, but the
proof-terms are created interactively, via tactics, where the typing
algorithm assists in filling the gaps in a proof. These are basically
`holes' in a proof term, but, different from Automath, a user never
actually observes the proof-term: it is created `under the hood'. In
Coq, one can write functional programs, prove them correct and
execute them in the system. A slightly modified version of the type
theory of Coq has been implemented in the Lean theorem prover
(\citealp{Mou}), with the specific aim of generating interest from the
mathematics community to formalize mathematical proofs.

\subsubsection{Agda}

The typed programming language Agda
(\citealp{Bov}) is based on dependent type theory and its underlying type
theory is Martin-L\"of Type Theory (\citealp{Nor}), which is close to
Coq's type theory. The focus is dependently typed programming, but it
is also a proof assistant, where one can formalize arbitrary notions
from mathematics and prove programs correct, just like Coq. The
interaction is quite different from Coq, because one edits proof-terms
directly. These are terms-with-holes, where the holes have types that
help the user to fill in the holes. The fact that one writes
proof-terms directly is closer to Automath, but, different from
Automath, the system gives a lot of interactive support for creating
these terms (i.e.\ filling the holes).

\end{document}